\documentclass[a4paper]{quantumarticle}

\pdfoutput=1

\usepackage[utf8]{inputenc}
\usepackage[english]{babel}
\usepackage[T1]{fontenc}
\usepackage{hyperref}

\usepackage[backend=bibtex]{biblatex}
\usepackage{graphicx}
\usepackage{float}

\usepackage{amsthm,amssymb,amsmath}

\theoremstyle{plain}

\theoremstyle{definition}
\newtheorem{definition}{Definition}

\theoremstyle{remark}

\begin{document}

\title{Experiment-friendly formulation of quantum backflow}

\author{Marek Miller}
\affiliation{School of Physical and Mathematical Sciences,
Nanyang Technological University, Singapore}
\affiliation{Centre of New Technologies, University of Warsaw, Poland}
\email{m.miller@cent.uw.edu.pl}

\author{Woo Chee Yuan}
\email{P160031@e.ntu.edu.sg}
\affiliation{School of Physical and Mathematical Sciences,
Nanyang Technological University, Singapore}

\author{Rainer Dumke}
\affiliation{School of Physical and Mathematical Sciences,
Nanyang Technological University, Singapore}
\affiliation{Centre for Quantum Technologies, National University of Singapore, Singapore}

\author{Tomasz Paterek}
\affiliation{School of Physical and Mathematical Sciences,
Nanyang Technological University, Singapore}
\affiliation{MajuLab, International Joint Research Unit UMI 3654, CNRS, Universite Cote d'Azur, Sorbonne Universite, National University of Singapore, Nanyang Technological University, Singapore}
\affiliation{Institute of Theoretical Physics and Astrophysics, 
	Faculty of Mathematics, Physics and Informatics,
	University of Gda\'nsk, Poland}

\begin{abstract}
Quantum backflow is usually understood as a quantum interference phenomenon where probability current of a quantum particle points in the opposite direction to particle's momentum.
Here, we quantify the amount of quantum backflow for arbitrary momentum distributions,
paving the way towards its experimental verification.
We give examples of backflow in gravitational and harmonic potential,
and discuss experimental procedures required for the demonstration using atomic gravimeters.
Such an experiment would show that the probability of finding a free falling particle above initial level could grow for suitably prepared quantum state with most momentum downwards.
\end{abstract}

\maketitle

\section{\label{sec:intro}Introduction}

A wave function of a quantum particle has physically observable characteristics that can be local or global.
The probability of finding the particle in a specific region of space or the probability current are examples of local characteristics, which can be determined given access to only small part of the wave function.
In contradistinction, the momentum is a property of the entire wave function,
e.g. requires the determination of the de Broglie wavelength.
Already at this level of generality,
it is clear that the probability current and the momentum of a quantum particle may behave very differently.

Quantum backflow (QB) is an interference effect that arises from this observation.  In order to understand the statement better, and to simplify the subsequent analysis, let us suppose that a particle in the one-dimensional space interacts with a potential that depends only on the particle's position.  Intuitively, we may think that if the momentum distribution concentrates within the positive half-line, the probability current, too, will be non-negative. 
It turns out that this is not the case for a suitably chosen quantum state.  In a sense, the probability flows `backwards'.  Hence the term ``quantum backflow''.

QB was first studied by Allcock in his work on the arrival time in quantum mechanics \cite{allcock1969time}
(see also Kijowski \cite{kijowski1974time} for an early discussion of QB).
Later, Bracken and Melloy \cite{bracken1994probability} gave a detailed analysis of QB as an eigenvalue problem.
The analysis was rigorously phrased in the mathematical language of 
integral operators on separable Hilbert space in \cite{penz2005new}.
In the same paper, a numerical approximation of the optimal QB state was given. 
See also \cite{halliwell2013quantum, yearsley2012analytical} for intresting case studies.
QB in systems interacting with linear potential was studied in
\cite{melloy1998velocity}.
Recently, there have been attempts at analysing QB
in the relativistic setting
\cite{melloy1998probability, su2018quantum, ashfaque2019relativistic},
in the setting of quantum particle decay
\cite{van2019decay,goussev2019equivalence},
as well as the attempts at describing quantum backflow in dissipative 
\cite{moussavi2020dissipative}
and many-particle systems \cite{barbier2020quantum}. 
The phenomenon was extended into phase space in Ref.~\cite{goussev2020probability} and systems suitable for observing QB in an experiment were discussed in Refs.~\cite{vasconcelos2020quantum,goussev2020quantum}.

Among others, Palmero \emph{et al}.~\cite{palmero2013detecting} proposed an experimental scheme that ``could lead to the observation of quantum backflow'' in $\phantom{}^7 \text{Li}$  Bose-Einstein condensate. 
To the best of our knowledge, however, QB has not yet been confirmed experimentally.
Note that in a recent experimental work by Eliezer~\emph{et al.}~ \cite{eliezer2018observation},
``optical backflow'' in transverse momentum of a beam of light was reported.
The present study, however, focuses on quantum backflow of individual non-relativistic quantum particles.
We extend the customary definition of QB to states with non-zero probability of measuring negative momentum. 
Hence, our approach should be applicable to realistic experimental situations.
We study systems exhibiting QB in gravitational field near the surface of the Earth, as well as interacting with harmonic potential.
We also comment on possible experimental verification of QB using atomic gravimeters.

\section{\label{sec:theory}Quantum backflow}

Our definition of quantum backflow conveys how necessary conditions for the probability current that follow from the classical equations of motion are no longer satisfied in the quantum regime.

Let us focus on the system of a lone particle in the one-dimensional space.  
We set the vertical direction with the $x$ axis pointing downwards.
Suppose the particle interacts with arbitrary potential $V(x)$. 
We examine its dynamics from classical and quantum points of view with initial conditions as similar as possible.
The quantum system at time $t$ is fully described by its wave function $\psi_t(x)$.
The  classical model requires simultaneous knowledge of position and momentum, whose precise estimation is famously forbidden by the uncertainty principle.
We therefore propose an operational approach in which distribution of position and momentum is estimated with finite precision, and
given by the probability density function
\begin{equation}
\label{eq:ftee}
f_{t}(x,p) = \left | \langle \phi | W(x,p) | \psi \rangle \right|^2,
\end{equation}
where $W(x,p)$ is the \mbox{Wigner-Moyal} transform 
of $\psi$ and $\phi$ (see e.g. Eq. (6.63) in \cite{de2006symplectic}):
\begin{multline}
\label{eq:wignermoyalTransform}
 \langle \phi | W(x,p) | \psi \rangle = \\
 = \frac{1}{2\pi \hbar}
 \int \limits_{-\infty}^{\infty}
    e^{-\frac{i}{\hbar} p y}  \phi^{*}(y- \frac{x}{2}) \, \psi(y + \frac{x}{2}) \, dy.
\end{multline}
The function $\phi$ represents the finite precision of the measurement
apparatus and, for example,
can be a Gaussian distribution, centred at zero,
with finite width $\sigma_\phi$.
The marginals of $f_t(x,p)$
agree with densities of position and momentum, derived from $\psi_t(x)$, "up to $\sigma_\phi^2$\,", i.e.:
\begin{align}
P_t(x = x_0) &=  \int \limits_{-\infty}^{\infty} f_t(x_0,p) dp  
    = (|\psi|^2 *|\phi|^2 )(x_0),  \\ 
P_t(p = p_0) &= \int \limits_{-\infty}^{\infty} f_t(x,p_0) dx  
    = (|\tilde{\psi}|^2 * |\tilde{\phi}|^2 )(p_0), 
\end{align}
where $\tilde{\psi}(p)$ and $\tilde{\phi}(p)$ are Fourier transforms of $\psi$ and $\phi$, respectively,
and the symbol '$*$' stands for convolution.  

In other words, the family of operators
$W(x,p)^{\dagger} |\phi \rangle \langle \phi | W(x,p)$,
defined on the phase space $(x,p) \in \mathbb{R}^2$,
is a positive operator-valued measure
that allows experimental estimation of
the joint probability distribution $f_t(x,p)$
of position and momentum in the state $\psi$,
with a finite precision given by a square-integrable function $\phi$.

Let us now derive probability currents in the quantum and classical models corresponding to the rate of change of probability of finding the particle above the level $x = a$.
In the quantum case,
this is a textbook exercise leading to the familiar formula
\begin{multline}
\label{eq:probcur}
j_t(a) =
- \frac{d}{dt} P_t(x \leq a) = \\
= \frac{\hbar}{m} \mathfrak{Im} \left(
\psi_t^{*}(a) \, \frac{d}{d x} \psi_t(x) \Bigr|_{x=a}
\right),
\end{multline}
where $\mathfrak{Im}$ stands for the imaginary part of a complex number.

Let us fix $t$ for now and 
consider a putative classical system, with the distribution of
position and momentum given by $f_{t}(x,p)$.
The distribution evolves for a short time $\tau$, $t \leq \tau \leq t + \Delta t$ according to the Hamilton equations of motion:
$\dot{x}(\tau) = p/m, \quad  \dot{p}(\tau) = - \frac{d V}{d x}$.
This implies that, unlike in the quantum case, the probability current of the classical system,
\begin{equation}
(j_{\textrm{cl}})_t(a) = - \frac{d}{d\tau} P_{\textrm{cl}} \left (x(\tau) \leq a \right ) \Big |_{\tau=t},
\end{equation}
must be bounded from below:
\begin{equation}
\label{eq:classBound}
(j_{\textrm{cl}})_t (a) \geq
\frac{1}{m} 
\int \limits_{-\infty}^{0}
     p \, f_{t} (a , p)
\,  dp
\end{equation}
(see Appendix for the detailed proof).
We can say that the probability of finding the classical particle above the line $x=a$ cannot grow faster than a certain quantity derived from the distribution of only negative momenta.  In particular, if $f_t(x,p) = 0$ for $p < 0$, we get that $(j_{\textrm{cl}})_t(a) \geq 0$, i.e. the direction of the momentum and the probability current coincide.
In this particular case, it leads to the usual definition of QB given by the following statements about a wave function of a quantum system in one-dimensional space~\cite{bostelmann2017quantum}:
a) $\tilde{\psi}_t$ contains only positive momenta;
b) there exists $a \in \mathbb{R}$, for which $j_{t}(a) < 0$.
To facilitate the analysis of QB for arbitrary states,
we say that QB is a situation where the inequality \eqref{eq:classBound} no longer holds.

\begin{definition}
	\label{DEF_QB}
The quantum backflow takes place at point $x=a$ and at time $t$, if
\begin{equation}
\label{eq:qbdef}
j_t(a) <
 \frac{1}{m} 
\int \limits_{-\infty}^{0}
     p  \, f_{t} (a , p)
\,  dp,
\end{equation}
where the probability current $j_t(a)$ is given by Eq. \eqref{eq:probcur}
and the function $f_t(x,p)$ by Eq. \eqref{eq:ftee}.
\end{definition}

Note that by taking a formal limit of the precision function converging to the Dirac delta distribution in the momentum representation,
$\tilde{\phi}(p) \rightarrow \delta(p)$, 
we can see from Eq.~\eqref{eq:wignermoyalTransform}
that $f_t(x,p) \rightarrow |\tilde{\psi}(p)|^2$,
at any point $x$.
Hence, in this limiting case, and for a wave function such that
$\tilde{\psi}(p) = 0$ for $p<0$, 
the inequality \eqref{eq:qbdef} reduces to 
the standard definition of QB, requiring only $j_t(a) <0$.

It is also worth mentioning that Eq.~\eqref{eq:qbdef} is a condition independent 
of whether the system undergoes quantum scattering.
For example, it is easy to see that an initial superposition of plain waves,
\begin{equation}
\psi(x) = A e^{ikx} + B e^{-ikx}, \quad x<0, 
\end{equation}
scattered back on a potential step, 
does not exhibit quantum backflow.
Unlike scattering, quantum backflow is an instantaneous property of
the particle's quantum state, namely its probability current exceeding a classical bound,
regardless of the underlying dynamics.

\begin{figure*}[t]
	\center
	\includegraphics[width=.45\linewidth]{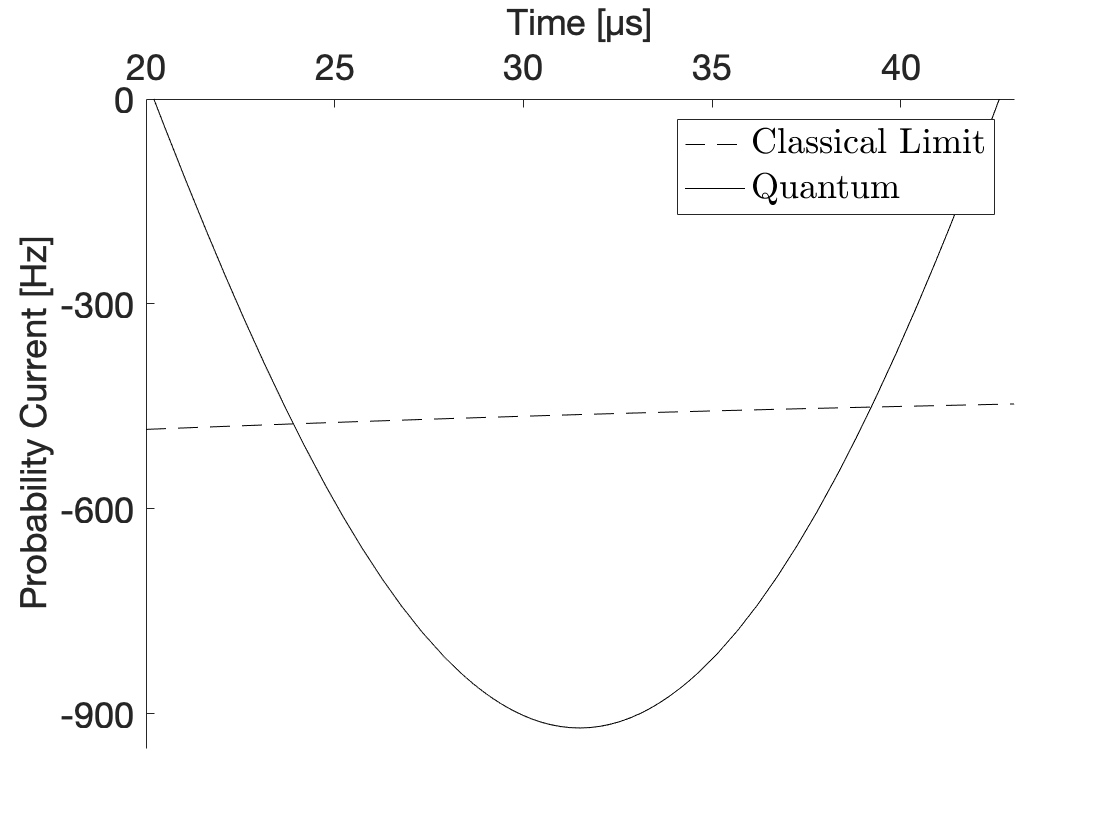}
	\includegraphics[width=.45\linewidth]{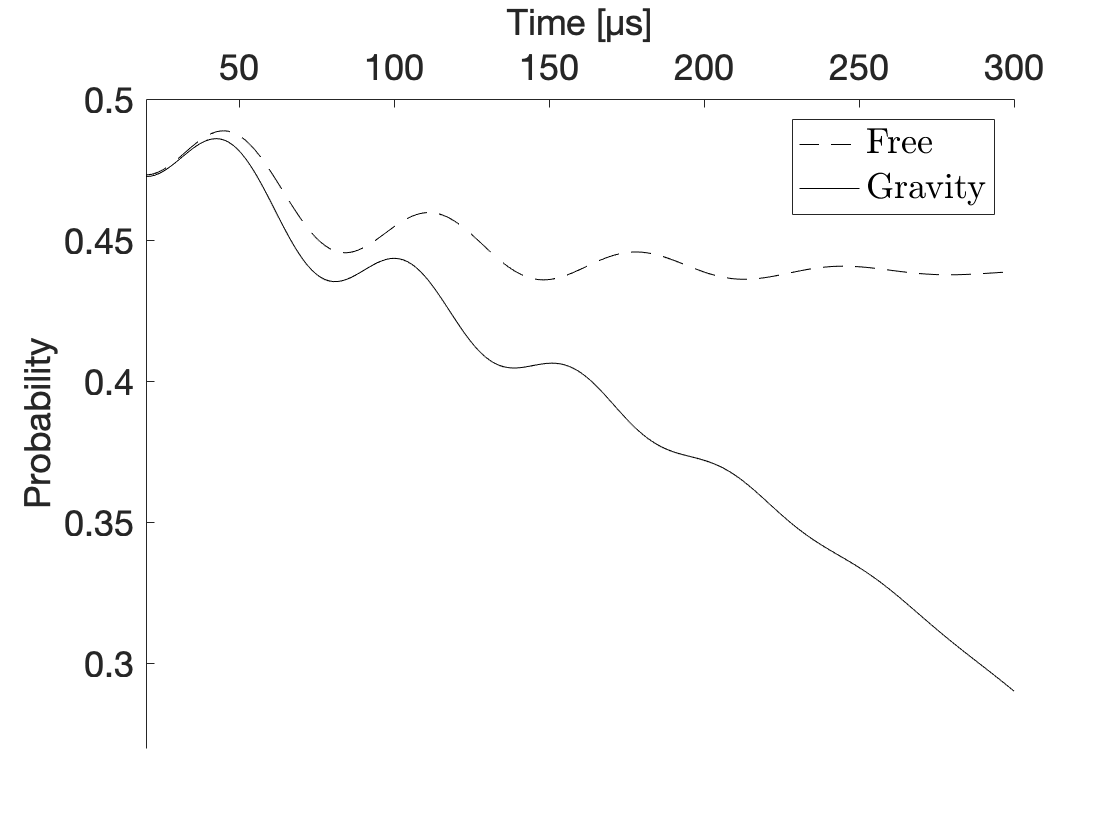}
	\caption{{\footnotesize Quantum backflow in gravity.
		The initial state is a superposition of two Gaussian wave functions with amplitudes given in~\eqref{eq:params}, describing Rubidium atom.
		Left panel:
		The dashed line gives the lower bound on the classical probability current.
		The solid line is the quantum probability current.
		QB takes place in the interval when the solid line is below the dashed line, see Eq.~\eqref{eq:qbdef}.
		Right panel:
		As a consequence,
		the probability $P(x<0)$ of finding the particle above the initial level of $x=0$ increases despite small contributions from negative momenta.
		Solid line gives the probability in the presence of the gravitational field. 
		For comparison, the dashed line represents the free particle.
	}}
	\label{fig:grav}
\end{figure*}

\section{\label{sec:predictions}Examples}

We give two concrete examples of QB according to Definition~\ref{DEF_QB}.
Both involve superposition of Gaussian states and therefore naturally contain contributions from negative momenta.
The first example is a particle in a linear potential as an approximation of the Newtonian gravity close to the Earth's surface.
The second example considers quadratic (harmonic) potential.

\subsection{Gravitational potential}

Suppose the particle interacts with the potential $V(x) = -mgx$, for $g\geq0$.
Recall that by our convention, the direction of the $x$ axis and the direction of the gravitational force coincide.
We choose the initial level $x = 0$ above the surface.
Of course, by putting $g=0$, we also cover the case of a free particle on the real line.

Consider the initial wave function, at time 
$t=0$, being a superposition of two Gaussian states centred at $x=0$, with the same spread $\sigma$ but different mean momenta.
If the corresponding quantum particle is free, the wave function at time $t$ reads:
\begin{multline}
\label{eq:freeSuperpos}
\psi_\mathrm{free}(x, t) = \sum_{n = 1}^{2} \frac{B_n}{\sqrt{4\sigma^2 + 2i\frac{h}{m} t}}
\times \\ \times
\exp\bigg(  \frac{i}{\hbar}p_n\big( x - \frac{p_n}{2m} t  \big ) - \frac{(x - \frac{p_n}{m} t )^2}{4\sigma^2 + 2i\frac{h}{m} t} \bigg),
\end{multline}
where $p_1, p_2$ are the mean values of momentum for each branch of the superposition and $B_1, B_2$ are the corresponding probability amplitudes.
In the presence of the linear potential $V(x)$,
the wave function accelerates as a result of the gravitational force, and at time $t$ it takes the form~\cite{vandegrift2000accelerating}:
\begin{equation}
\psi(x,t) =
e^{-\frac{i}{\hbar}(- m g t x + \frac{1}{6}mg^2t^3)} \, \psi_{\mathrm{free}}( x-\frac{1}{2}gt^2, t).
\end{equation}

Let us now choose realistic values of the parameters
(see next section for further discussion of a possible experiment).
E.g., let us take the Rubidium atom of mass $m \approx 1.4 \times 10^{-25}$ kg, described by the wave function with $\sigma = 1 \mu$m.
The mean momenta are set to $p_1 = 0$ and $p_2 = 2 \hbar k$, where the wave number $k = 2 \pi / \lambda$ with $\lambda = 780$ nm. The following amplitudes
\begin{equation}
B_1 \approx 1.18 \times 10^{-3}, \ B_2 \approx 4.42  \times 10^{-4},
\label{eq:params}
\end{equation}
numerically optimise the effect of QB.
The precision function is set to be a Gaussian with standard deviation $\sigma_\phi = 0.1 \mu$m. 

In Fig.~\ref{fig:grav}, 
we show
the probability current and the overall probability of finding the particle above the initial line $x = 0$ as functions of time. 
We set $g = 9.8 \, \text{m} \cdot \text{s}^{-2}$.
It is clear that for $t$
such that approximately:
$23.7 \mu$s $\leq t \leq 39 \mu$s,
the probability current
satisfies the inequality \eqref{eq:qbdef}, i.e. QB takes place.
During that time, 
the numerical value of the integral 
$\int_{-\infty}^0 |\tilde{\psi}(p)|^2 \, dp$, i.e. the contribution of ``negative momenta'' to the backflow state, ranges from $0.23$ to $0.13$.
Here, our approach allows us to separate the contribution of the negative momenta from the genuine quantum backflow.
Despite the particle's free fall, the quantum probability of finding the particle above the initial level unmistakably \emph{increases}.

\subsection{Harmonic potential}

\begin{figure}
	\center
	\includegraphics[width=.95\linewidth]{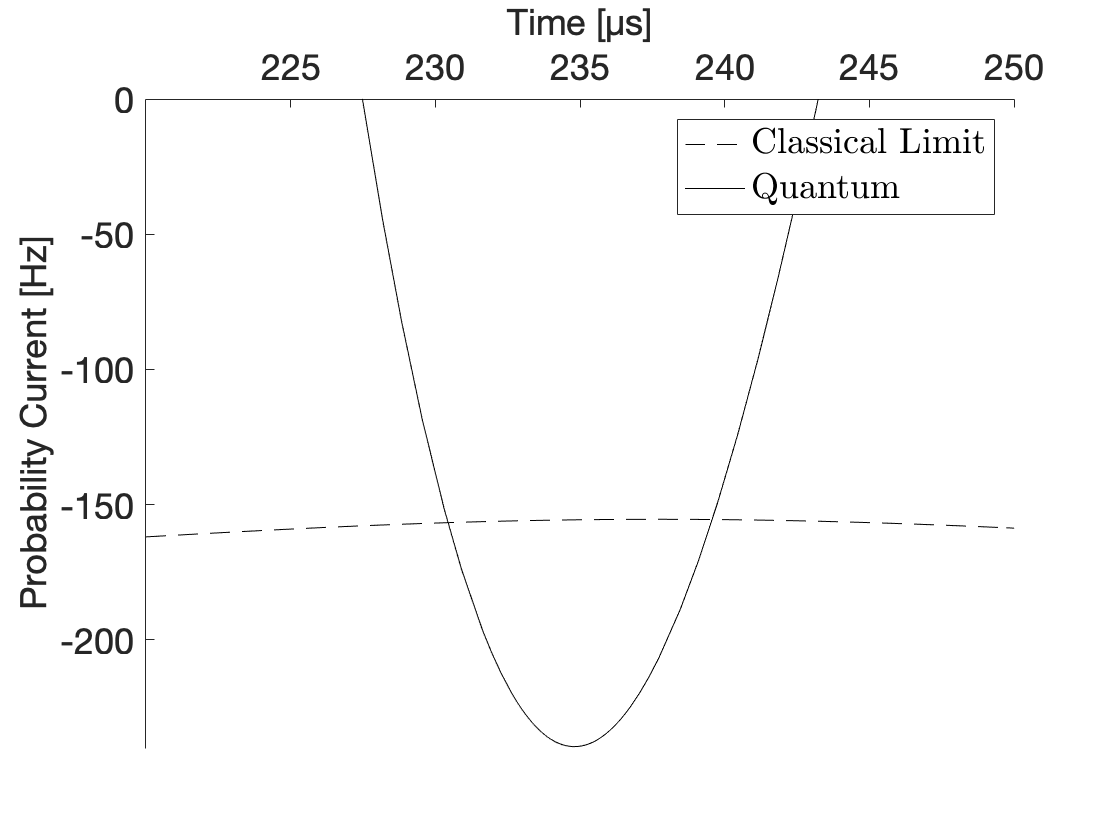}
	\caption{
		Quantum backflow in a harmonic trap.
		Rb atom in the initial superposition of coherent states with parameters given in the main text gives rise to the quantum probability current (solid line) which is below the classical lower bound (dashed line).
		This signifies QB for approximately $(230, 239) \mu$s. Probability current was evaluated at $x = \sqrt{\frac{2\hbar}{m\omega}} \cos{(0.55\pi)}$.}
	\label{fig:cohere}
\end{figure}

Our second example is a particle in quadratic potential, due to its wide applicability.
We again consider Rb atom, but this time in a harmonic trap with frequency $\nu = 10$ kHz \cite{Beams2004}.
It is generally known that coherent states of a quantum harmonic oscillator take the form of a Gaussian packet in position representation \cite{klauder1985coherent}.  
Hence, we take as an initial state the superposition of two coherent states $|\psi \rangle = a |\alpha \rangle + a |\beta \rangle$. Numerical optimisation of QB leads to the following parameters: $a \approx 0.73$, $\alpha = e^{i (0.9\pi - \omega t)}$, $\beta = 9e^{i (0.55\pi - \omega t)}$, where $\omega = 2 \pi \nu$.
Fig.~\ref{fig:cohere} shows the corresponding classical bound on the probability current and the quantum prediction for QB.

\section{Experimental proposal}

The approach proposed here is based on the measurement of probability current and generalised joint measurement (with finite precision) of position and momentum.
The latter determines the right-hand side of Eq.~\eqref{eq:qbdef} and the former its left-hand side.
Probability current is a convenient theoretical quantity in this problem because in the derivation of \eqref{eq:qbdef} Hamilton's equations reduce the classical bound to the integral over momentum only.
In more practical terms, the measurement of the probability current requires estimation of particle's position at two nearby times and division by the corresponding time interval.
This can be simplified to only one estimation of particle's position by using the bound of Eq.~\eqref{APP_EQ_P_BOUND} in the Appendix.

The essence of the present proposal is to establish the initial joint probability distribution of position and momentum that is used to compute the classical bound.
Since this might be experimentally quite demanding, we offer one more route to the verification of QB, based on independent measurements of initial position and momentum distributions.
Such measurements do not give any information about possible position-momentum correlations.
Accordingly, we propose to consider all joint distributions $f_t(x,p)$ compatible with these marginals and take the worst case upper bound in Eq.~\eqref{APP_EQ_P_BOUND}.

Finally, a concrete setup capable of measuring QB could be constructed using atomic gravimeters.
High precision estimation of position distribution has been realised with electron~\cite{SEM_ATOM} as well as optical microscopy~\cite{OPT1,OPT2} of the atomic cloud;
and momentum distribution can be established from the time-of-flight measurements~\cite{CORNELL96}.
Depending on the way the atoms are trapped these setups offer initial width of the wave function in the range from micrometers to millimetres.
We have verified numerically that QB is present for all these spreads with the order of magnitude of the bound and the quantum probability current scaling like inverse square of the width.

\section{\label{sec:discus}Conclusions}

We presented an analysis of QB for states with admissible negative momentum. 
Unlike the standard definition of QB, Def. \ref{DEF_QB} is applicable not only to wave functions with positive support in momentum representation.  
Broadly speaking, our idea is to compare the classical evolution of the joint probability distribution of position and momentum (known with finite precision) with the quantum dynamics.  The classical system would evolve for an infinitesimally short time according to Hamilton's equations of motion, starting with the same initial conditions.  We defined QB as a situation when the probability current exceeds what might be possible for its classical counterpart.

In particular, we showed that a relatively easy to prepare superposition of two Gaussian packets exhibits QB during free-fall in the uniform gravitational field near the Earth's surface.
Of course, the Ehrenfest theorem guarantees that the average position of the quantum particle follows the classical trajectory.  Nevertheless, the probability of locating the particle above the initial level displays `antigravitational' quantum backflow.

\section{Acknowledgments}
We would like to warmly thank Referees for their insightful comments and questions. 

This work is supported by the Polish National Agency for Academic Exchange NAWA Project No. PPN/PPO/2018/1/00007/U/00001.

\appendix
\newpage
\section{Appendix}
Here, we give a proof of the inequality \eqref{eq:classBound}.

Once again, we consider the distribution of
position and momentum given by $f_{t}(x,p)$
that evolves classically for a brief time $\tau$, $t \leq \tau \leq t + \Delta t$.
If our system were initially at point $x$, then
$x(\tau) = x + \frac{p}{m} (\tau-t)$.
The probability $P(x(\tau) \leq a)$ of finding the particle in the region
``above'' the line $x=a$ would be
\begin{eqnarray}
P(x(\tau) \leq a) & = & P(x \leq a - \frac{p}{m} (\tau - t) ) \nonumber \\
& = & \int \limits_{- \infty}^{a - \frac{p}{m} (\tau-t)} \int \limits_{- \infty}^{\infty} f_t(x,p) \, dp \, dx \label{APP_EQ_LIMIT} \\
& = & \int \limits_{- \infty}^{a} \int \limits_{- \infty}^{\infty} f_t(x - \frac{p}{m} (\tau-t),p) \, dp \, dx. \nonumber
\end{eqnarray}

Now, the probability current of the classical system,
\begin{equation}
(j_{\textrm{cl}})_t(a) = - \frac{d}{d\tau} P \left (x(\tau) \leq a \right ) \Big |_{\tau=t},
\end{equation}
can be expressed 
in terms of the probability density function $f_t(x,p)$:
\begin{eqnarray}
(j_{\textrm{cl}})_t(a) & = & - \frac{d}{d\tau} 
\int \limits_{- \infty}^{a} \int \limits_{- \infty}^{\infty} f_t(x - \frac{p}{m} (\tau-t),p) \, dp \, dx \Big |_{\tau=t} \nonumber \\
& = & \int \limits_{-\infty}^{a} 
\int \limits_{-\infty}^{\infty} 
    \frac{p}{m} \frac{\partial}{\partial x} \, f_{t} (x , p) 
\, dp \, dx \nonumber \\
& = & \frac{1}{m}  
\int \limits_{-\infty}^{\infty} 
     p f_{t} (a , p) 
\, dp ,
\end{eqnarray}
where the last equality is a consequence of the fact that 
$\lim_{x\rightarrow\pm \infty} f_t(x,p) = 0$.
This immediately yields the inequality:
\begin{equation}
(j_{\textrm{cl}})_t (a) \geq
 \frac{1}{m} 
\int \limits_{-\infty}^{0}
     p  f_{t} (a , p) \,  dp.
\end{equation}

From a practical point of view, in order to save on measurements, it would be more desirable to derive the classical upper bound for the probability of finding the particle above the initial level instead of the lower bound on the probability current.
For this we use Eq.~\eqref{APP_EQ_LIMIT} and note the following inequality
\begin{align}
	P(x(\tau) \leq a) 
	& \leq  \int \limits_{- \infty}^{a - \frac{p}{m} (\tau-t)} \int \limits_{- \infty}^{0} f_t(x,p) \, dp \, dx \nonumber \\
	& +  \int \limits_{- \infty}^{a} \int \limits_{0}^{\infty} f_t(x,p) \, dp \, dx,
\end{align}

which is simply a consequence of non-negativity of $f_t(x,p)$.
Hence,
\begin{align}
\label{APP_EQ_P_BOUND}
	P(x(\tau) \leq a)  & \leq	P(x \leq a)    \nonumber \\ 
	&+ \int_{a}^{a - \frac{p}{m} (\tau-t)} \int \limits_{- \infty}^{0} f_t(x,p) \, dp \, dx.
\end{align}

\printbibliography

@article{ashfaque2019relativistic,
  title={Relativistic quantum backflow},
  author={Ashfaque, J. and Lynch, J. and Strange, P.},
  journal={Physica Scripta},
  year={2019},
  publisher={IOP Publishing},
  doi={https://doi.org/10.1088/1402-4896/ab265c}
}

@article{bostelmann2017quantum,
  title={Quantum backflow and scattering},
  author={Bostelmann, H. and Cadamuro, D. and Lechner, Ga.f},
  journal={Physical Review A},
  volume={96},
  number={1},
  pages={012112},
  year={2017},
  publisher={APS},
  doi={https://doi.org/10.1103/PhysRevA.96.012112}
}

@article{palmero2013detecting,
  title={Detecting quantum backflow by the density of a Bose-Einstein condensate},
  author={Palmero, M. and Torrontegui, E. and Muga, J.G. and Modugno, M.},
  journal={Physical Review A},
  volume={87},
  number={5},
  pages={053618},
  year={2013},
  publisher={APS},
  doi={https://doi.org/10.1103/PhysRevA.87.053618}
}

@article{yearsley2012analytical,
  title={Analytical examples, measurement models, and classical limit of quantum backflow},
  author={Yearsley, J.M. and Halliwell, J.J. and Hartshorn, R. and Whitby, A.},
  journal={Physical Review A},
  volume={86},
  number={4},
  pages={042116},
  year={2012},
  publisher={APS},
  doi={https://doi.org/10.1103/PhysRevA.86.042116}
}

@article{penz2005new,
  title={A new approach to quantum backflow},
  author={Penz, M. and Gr{\"u}bl, Ge. and Kreidl, S. and Wagner, P.},
  journal={Journal of Physics A: Mathematical and General},
  volume={39},
  number={2},
  pages={423},
  year={2005},
  publisher={IOP Publishing},
  doi={https://doi.org/10.1088/0305-4470/39/2/012}
}

@article{bracken1994probability,
  title={Probability backflow and a new dimensionless quantum number},
  author={Bracken, A.J. and Melloy, G.F.},
  journal={Journal of Physics A: Mathematical and General},
  volume={27},
  number={6},
  pages={2197},
  year={1994},
  publisher={IOP Publishing},
  doi={https://doi.org/10.1088/0305-4470/27/6/040}
}

@article{allcock1969time,
  title={The time of arrival in quantum mechanics I. Formal considerations},
  author={Allcock, Go.R.},
  journal={Annals of Physics},
  volume={53},
  number={2},
  pages={253--285},
  year={1969},
  publisher={Elsevier},
  doi={https://doi.org/10.1016/0003-4916(69)90251-6}
}

@article{eliezer2018observation,
  title={Observation of Optical Backflow},
  author={Eliezer, Y. and Zacharias, T. and Bahabad, A.},
  journal={Optica},
 volume={7},
  pages={72},
  year={2020},
  doi={https://doi.org/10.1364/OPTICA.371494}
}

@article{su2018quantum,
  title={Quantum backflow in solutions to the Dirac equation of the spin-1 2 free particle},
  author={Su, H.-Y. and Chen, J.-L.},
  journal={Modern Physics Letters A},
  volume={33},
  number={32},
  pages={1850186},
  year={2018},
  publisher={World Scientific},
  doi={https://doi.org/10.1142/S0217732318501869}
}

@article{melloy1998velocity,
  title={The velocity of probability transport in quantum mechanics},
  author={Melloy, G.F. and Bracken, A.J.},
  journal={Annals of Physics},
  volume={7},
  number={7-8},
  pages={726-731},
  year={1998},
  publisher={},
  doi={https://doi.org/10.1002/(SICI)1521-3889(199812)7:7/8%3C726::AID-ANDP726%3E3.0.CO;2-P}
}

@article{halliwell2013quantum,
  title={Quantum backflow states from eigenstates of the regularized current operator},
  author={Halliwell, J.J. and Gillman, E. and Lennon, O. and Patel, M. and Ramirez, I.},
  journal={Journal of Physics A: Mathematical and Theoretical},
  volume={46},
  number={47},
  pages={475303},
  year={2013},
  publisher={IOP Publishing},
  doi={https://doi.org/10.1088/1751-8113/46/47/475303}
}

@article{moussavi2020dissipative,
  title={Dissipative quantum backflow},
  author={Mousavi, S..V and Miret-Art{\'e}s, S.},
  journal={The European Physical Journal Plus},
  volume={135},
  number={3},
  pages={1--18},
  year={2020},
  publisher={Springer},
  doi={https://doi.org/10.1140/epjp/s13360-020-00336-5}
}

@book{de2006symplectic,
  title={Symplectic geometry and quantum mechanics},
  author={De Gosson, M.A.},
  volume={166},
  year={2006},
  publisher={Springer Science \& Business Media},
  doi={https://doi.org/10.1007/3-7643-7575-2}
}

@article{vandegrift2000accelerating,
  title={Accelerating wave packet solution to Schr{\"o}dinger’s equation},
  author={Vandegrift, G.},
  journal={American Journal of Physics},
  volume={68},
  number={6},
  pages={576--577},
  year={2000},
  publisher={AAPT},
  doi={https://doi.org/10.1119/1.19492}
}

@book{klauder1985coherent,
  title={Coherent States: Applications in Physics and Mathematical Physics},
  author={Klauder, J.R. and Skagerstam, B.-S.},
  year={1985},
  publisher={World Scientific, Singapore},
  doi={https://doi.org/10.1142/9789814415118_0007}
}

@article{Beams2004,
	doi = {https://doi.org/10.1088/0953-4075/37/22/014},
	year = 2004,
	publisher = {{IOP} Publishing},
	volume = {37},
	number = {22},
	pages = {4561--4570},
	author = {Beams, T.J. and Peach, G. and Whittingham, I.B.},
	title = {Ultracold atomic collisions in tight harmonic traps: perturbation theory, ionization losses and application to metastable helium atoms},
	journal = {Journal of Physics B: Atomic, Molecular and Optical Physics}
}

@article{melloy1998probability,
  title={Probability backflow for a Dirac particle},
  author={Melloy, G.F. and Bracken, A.J.},
  journal={Foundations of Physics},
  volume={28},
  number={3},
  pages={505--514},
  year={1998},
  publisher={Springer},
  doi={https://doi.org/10.1023/A:1018724313788}
}

@article{van2019decay,
  title={Decay of a quasistable quantum system and quantum backflow},
  author={van Dijk, W. and Toyama, F.M.},
  journal={Physical Review A},
  volume={100},
  number={5},
  pages={052101},
  year={2019},
  publisher={APS},
  doi={https://doi.org/10.1103/PhysRevA.100.052101}
}

@article{goussev2020probability,
  title = {Probability backflow for correlated quantum states},
  author = {Goussev, A.},
  journal = {Physical Review Research},
  volume = {2},
    @issue = {3},
  pages = {033206},
  numpages = {10},
  year = {2020},
  publisher = {American Physical Society},
  doi = {https://doi.org/10.1103/PhysRevResearch.2.033206}
}

@article{goussev2019equivalence,
  title={Equivalence between quantum backflow and classically forbidden probability flow in a diffraction-in-time problem},
  author={Goussev, A.},
  journal={Physical Review A},
  volume={99},
  number={4},
  pages={043626},
  year={2019},
  publisher={APS},
  doi={https://doi.org/10.1103/PhysRevA.99.043626}
}

@article{barbier2020quantum,
  title = {Quantum backflow for many-particle systems},
  author = {Barbier, M.},
  journal = {Physical Review A},
  volume = {102},
  pages = {023334},
  numpages = {11},
  year = {2020},
  publisher = {American Physical Society},
  doi = {https://10.1103/PhysRevA.102.023334}
}

@article{kijowski1974time,
  title={On the time operator in quantum mechanics and the Heisenberg uncertainty relation for energy and time},
  author={Kijowski, J.},
  journal={Reports on Mathematical Physics},
  volume={6},
  number={3},
  pages={361--386},
  year={1974},
  publisher={Elsevier},
  doi={https://doi.org/10.1016/S0034-4877(74)80004-2}
}

@article{vasconcelos2020quantum,
  title={Quantum backflow in the presence of a purely transmitting defect},
  author={de Vasconcelos Jr, A.H.},
  journal={arXiv preprint arXiv:2007.07393},
  year={2020}
}

@article{goussev2020quantum,
  title={Quantum backflow in a ring},
  author={Goussev, A.},
  journal={arXiv preprint arXiv:2008.08022},
  year={2020}
}

@article{SEM_ATOM,
  title={High-resolution scanning electron microscopy of an ultracold quantum gas},
  author={Gericke, T. and W\"urtz, P. and Reitz, D. and Langen, T. and Ott, H.},
  journal={Nat. Phys.},
 volume={4},
  pages={949–953},
  year={2008},
  doi={https://doi.org/10.1038/nphys1102}
}

@article{OPT1,
  title={Super-resolution microscopy of single atoms in optical lattices},
  author={Alberti, A. and Robens, C. and Alt, W. and Brakhane, S. and Karski, M. and Reimann, R. and Widera, A. and Meschede, D.},
  journal={New J. Phys.},
 volume={18},
  pages={053010},
  year={2016},
  doi={https://doi.org/10.1088/1367-2630/18/5/053010}
}

@article{OPT2,
  title={Superresolution Microscopy of Cold Atoms in an Optical Lattice},
  author={McDonald, M. and Trisnadi, J. and Yao, K.-X. and Chin, C.},
  journal={Phys. Rev. X},
 volume={9},
  pages={021001},
  year={2019},
  doi={https://doi.org/10.1103/PhysRevX.9.021001}
}

@article{CORNELL96,
  title={Bose-Einstein Condensation in a Dilute Gas: Measurement of Energy and Ground-State Occupation},
  author={Ensher, J.R. and Jin, D.S. and Matthews, M.R. and Wieman, C.E. and Cornell, E.A.},
  journal={Phys. Rev. lett.},
 volume={77},
  pages={4984},
  year={1996},
  doi={https://doi.org/10.1103/PhysRevLett.77.4984}
}

\end{document}